\documentclass[11pt]{article}
\usepackage{graphicx}
\usepackage{amsmath}
\allowdisplaybreaks[2]
\newcommand{\bG}{\ensuremath{\boldsymbol{G}}}

\newcommand{\bR}{\ensuremath{\boldsymbol{R}}}
\newcommand{\bC}{\ensuremath{\boldsymbol{C}}}
\newcommand{\bD}{\ensuremath{\boldsymbol{D}}}
\newcommand{\bB}{\ensuremath{\boldsymbol{B}}}
\newcommand{\bg}{\ensuremath{\boldsymbol{g}}}
\newcommand{\bc}{\ensuremath{\boldsymbol{c}}}

\newcommand{\bxi}{\ensuremath{\boldsymbol{\xi}}}
\newcommand{\bpsi}{\ensuremath{\boldsymbol{\psi}}}
\newcommand{\bsigma}{\ensuremath{\boldsymbol{\sigma}}}
\newcommand{\bfeta}{\ensuremath{\boldsymbol{\eta}}}
\newcommand{\bgamma}{\ensuremath{\boldsymbol{\gamma}}}
\newcommand{\nn}{\ensuremath{\eta}}
\newcommand{\tn}{\ensuremath{\phi}}
\newcommand{\av}[1]{\ensuremath{\left\langle #1 \right\rangle}}
\newcommand{\avs}[1]{\ensuremath{\langle\hspace{-2pt}\langle 
                                         #1
				 \rangle\hspace{-2pt}\rangle
				      }}				      
\newcommand{\avg}[1]{\ensuremath{\Big\langle\!\!\Big\langle 
                                         #1
				 \Big\rangle\!\!\Big\rangle
				      }}				      
\newcommand{\avgl}[1]{\ensuremath{\Big\langle\!\!\Big\langle 
                                         #1
				      }}				      
\newcommand{\avgr}[1]{\ensuremath{
                                         #1
				 \Big\rangle\!\!\Big\rangle
				      }}				      
\newcommand{\avgg}[1]{\ensuremath{\Bigg\langle\!\!\!\Bigg\langle 
                                         #1
				 \Bigg\rangle\!\!\!\Bigg\rangle
				      }}				      
\newcommand{\avv}[2]{\ensuremath{
     \left\langle
        #1
     \right\rangle_{#2}
     }}
\newcommand{\ma}[1]{\ensuremath{\hat{#1}}}
\newcommand{\one}{{ \boldsymbol{I}}}

\newcommand{\order}[1]{{\cal O}(#1)}
\newcommand{\distnorm}{\ensuremath{{\cal N}}}
\newcommand{\sgn}{\mbox{sgn}}

\newcommand{\gauss}[1]{{\cal D} #1}

\begin{document}

\title
   {\bf The signal-to-noise analysis of the  Little-Hopfield model
       revisited}
\author{D. Boll\'e, J. Busquets Blanco  and T. Verbeiren}
\date{}
\maketitle

\begin{abstract}

Using the generating functional analysis an exact recursion relation is
derived for the time evolution of the effective local field of the fully
connected Little-Hopfield model. It is shown that, by leaving out the
feedback correlations arising from earlier times in this effective
dynamics, one precisely finds the recursion relations usually employed
in the signal-to-noise approach.
The consequences of this approximation as well as the physics behind it
are discussed. In particular, it is pointed out why it is hard to
notice the effects, especially for model parameters corresponding to
retrieval. Numerical simulations confirm these findings.
The signal-to-noise analysis is then extended to include all correlations,
making it a full theory for dynamics at the level of the  generating
functional analysis.
The results are applied to the frequently employed  extremely diluted
(a)symmetric architectures  and to sequence processing networks.
\end{abstract}


\section{Introduction}

During the last number of years, the treatment of dynamics using generating
functional techniques (GFA) has received a lot of attention in the field
of statistical mechanics of disordered systems, in particular neural
networks (see, e.g., \cite{GCS03}--\cite{KO02} and references therein).
Such a treatment allows for the exact solution of the dynamics and finds all
relevant physical order parameters at any time step via the derivatives of
a generating functional \cite{MSR73}--\cite{HBFKS89}.
An alternative method to study dynamics of neural networks is the so called
signal-to-noise analysis (SNA) or statistical neurodynamics (see, e.g.,
\cite{AM88}--\cite{B03})
where the idea is to start from a splitting of the local field into a
signal part originating from the pattern to be retrieved and a noise part
arising from the other patterns. The differences in the existing versions
of this approach consist out of the different treatments of this noise
term, ranging from the assumption that it is Gaussian with various
approximations for its variance  \cite{AM88,O95,K85} to a
supposedly exact treatment \cite{PZ91,PZ91b}.

In two recent papers some comparisons have been made between the two methods
for a sequence processing network \cite{KO02} respectively the fully
connected Blume-Emery-Griffiths network \cite{BBSV03}. For the first
system, surprisingly, the order parameter equations obtained through the
exact GFA solution are shown to be completely equivalent to those of
statistical neurodynamics, known to be an approximation that assumes
Gaussian noise. These theoretical results are  verified by computer
simulations. We recall that this system contains no feedback correlations.
For the second system that is fully connected and, hence, does contain
feedback correlations, it has been shown that the results of the GFA and SNA
coincide up to the third time step. Some numerical experiments then indicated
that they may differ for further time steps, certainly for those parameters
of the system corresponding to spin-glass behaviour.

The idea of the present work is to perform a systematic analytical study
of the relationship between both techniques using the fully connected
Little-Hopfield model. In order
to do so, we take the GFA one step
further by deriving a recursion relation for the effective local field.
To our knowledge such a recursion relation has not yet been reported
upon in the literature. Furthermore, it is precisely this relation that
we use as basis for studying the correspondence between both methods.

For the fully connected model we show that the SNA as it has been
applied up to now in fact approximates the exact dynamics because it
forgets about a part of the  correlations. We discuss the physics
behind such a short-memory approximation and explain why it leads to
very good results in the case of retrieval. Moreover, we show how to apply
the SNA correctly leading to a complete equivalence with the GFA. The results
obtained are applied to other architectures and to sequence processing
networks.

The  paper is organized as follows. In Section 2 we recall the fully
connected Little-Hopfield model and discuss the SNA approach to solve its
dynamics. Section 3 shortly reviews the GFA and  derives recursion relations
for the effective local field in this framework.  Section 4 introduces the
short-memory approximation and shows how this reduces the results of the GFA
to those of the SNA. It also explains the physics behind this
approximation and discusses some numerical results. Section 5  presents
a scheme how to apply exactly the SNA. In section 6 we apply our
findings to the extremely diluted symmetric and asymmetric architectures
and to sequence processing models. Finally, Section 7 contains some
concluding remarks.

\section{Signal-to-noise analysis of the Little-Hopfield model}

\subsection{The Little-Hopfield model}

By now, the Little-Hopfield model is a standard model for associative
memory and can be found in many textbooks (see, e.g, \cite{Nbook}).
Consider a system of $N$ Ising spins $\sigma_i$, $i=1, \dots, N$.  We
want to store $p=\alpha N$ patterns
${\bxi}^\mu = \{\xi^\mu_1, \dots, \xi^\mu_N\}$, $\mu=1, \dots, p$,
independent and identically distributed
with respect to $i$ and $\mu$.
The local field in neuron $i$ is defined by
\begin{equation}
h_i(t) = \sum_{j=1}^N J_{ij} \, \sigma_j(t)\, ,
\end{equation}
with the couplings given by the Hebb rule
\begin{equation}
J_{ij} = \frac{1}{N} \sum_{\mu=1}^p \xi_i^\mu \xi_j^\mu \,,
       \quad J_{ii}=0 \ .
\end{equation}
All neurons are updated in parallel according to the Glauber dynamics
\begin{equation} \label{eq:dyn:glauber}
\mbox{Prob}[\sigma_i(t+1) = s|h_i(t)] =
         \frac{e^{\beta s h_i(t)}}{2 \cosh(\beta h_i(t))} \ ,
	 \quad s=\pm 1 \, ,
\end{equation}
which becomes, in the limit $\beta= 1/T\to\infty$, equivalent to the
gain function formulation
\begin{equation}
           \sigma_i(t+1) = \sgn(h_i(t)) \ .
\end{equation}
In general, we  write
\begin{equation}
    \label{eq:dyn:det:gen}
         \sigma_i(t+1) = g(h_i(t)) \, .
\end{equation}

The long time behaviour of the system is governed by the Hamiltonian
\cite{P84}
\begin{equation}
H = -\frac{1}{\beta} \sum_i \ln[\cosh(\beta h_i(t))] \,.
\end{equation}
The thermodynamic and retrieval properties are visualized in the
 $(T-\alpha)$ phase diagram \cite{FK88}. There is a
transition curve starting at $T=0, \alpha=0.138$ and ending at
$T=1,\alpha=0$ beyond which the system does not retrieve any patterns
anymore and behaves like a spin-glass.  For higher temperatures the system
undergoes a transition to the paramagnetic phase.

The parallel dynamics of this model using both the SNA and GFA and,
especially, the comparison between these approaches is the subject of the
following sections.

\subsection{Signal-to-noise analysis}

In order to keep this paper self-contained we shortly review the
signal-to-noise analysis.
We assume that the system has an initial finite overlap with one of
the patterns, say the first one, which we call condensed.  The other
patterns (non-condensed) act as noise, making it harder for the
network to retrieve the condensed pattern.

We first focus on zero temperature.
The key idea is to separate the signal from the noise in the local field
\begin{equation} \label{eq:h:signal-noise}
h_i(t)  = \xi_i^1 \frac{1}{N} \sum_{j} \xi_j^1 \sigma_j(t)
         + \frac{1}{N} \sum_{\mu\ne1} \xi^\mu_i \sum_{j}
          \xi_j^\mu \sigma_j(t)
         - \alpha \sigma_i(t)
\end{equation}
whereby it is technically convenient to include the self-interaction and
subtract it again leading to the term $\alpha\sigma_i(t)$.  Quantities that
have a site-index, or an index $N$ are quantities for a finite system.
We define the overlap, respectively the residual overlap  by
\begin{equation}
     \label{eq:op:def}
     m_N(t)= \frac{1}{N}\sum_{j} \xi_j^1 \sigma_j(t)\, ,
  \qquad
    r_{N}^\mu(t)= \frac{1}{\sqrt{N}} \sum_{j} \xi_j^\mu \sigma_j(t)
    \, ,\quad \mu >1 \,.
\end{equation}
The local field (\ref{eq:h:signal-noise}) then becomes
\begin{equation}
   \label{eq:h:signal-noise:1}
    h_i(t)  = \xi_i^1 m_N(t)
         + \frac{1}{\sqrt{N}} \sum_{\mu\ne1} \xi^\mu_i r^\mu_N(t)
         - \alpha \sigma_i(t) \, .
\end{equation}

Our aim is to determine the form of the local field in the
thermodynamic limit $N \to \infty$. To the signal term we apply the law of
large numbers (LLN)
\begin{equation}
    \lim_{N \to \infty} m_N(t) \stackrel{{\mathcal Pr}}{=} m(t)
    \label{eq:blabla}
\end{equation}
where the convergence is in probability.
Taking a closer look at the second term in (\ref{eq:h:signal-noise:1}),
we could apply the central limit theorem (CLT) if all the terms in the sum
would be independent. This, of course, depends on the architecture and it
is  true, e.g.,  for asymmetrically extremely diluted models \cite{DGZ87,
KZ91}. In general, there
are non-trivial correlations between the terms and the different
implementations of the SNA mentioned before treat these correlations in
a different way.

For the fully connected architecture at hand, the most naive and simple
approach \cite{K85} is to keep the assumption that all terms in the
sum are uncorrelated, and that one can simply apply the CLT theorem to the
second term in (\ref{eq:h:signal-noise:1}). The local field  becomes
normally distributed with mean $\xi_i^1m(t)$ and variance $\alpha$.
This approach has then been refined in the theory of statistical
neurodynamics \cite{AM88}. There one also assumes that the noise part of the
local field is normally distributed but one calculates explicitly its
variance starting from its definition and taking into account part of the
correlations between the embedded pattern and neuron states in the dynamics.
This results in a more complex structure of the variance of the local field.
For more technical details we refer to \cite{AM88,Nbook}.
Detailed simulations show that the assumption of Gaussian
noise is approximately valid and close to reality as long as retrieval is
successful. In particular, it succeeds in explaining qualitatively the
dynamical behaviour of retrieval in the associative memory model.
However, in addition to a different critical capacity at $T=0$, the basin
of attraction calculated in this scheme is larger than the one
obtained by computer simulations \cite{NO93}.
This is attributed to the fact that the correlations of the local fields
at successive time steps are neglected. More specific, the average over
the Gaussian distribution of the local field at time $t+1$ is taken
independently of the average of the neuron state at time $t$, which
clearly depends on the local field at time $t-1$.

Later on these correlations between the local field at successive time
steps (and therefore also correlations between neuron states at different
time steps) have  been taken into account. In this treatment one
rewrites the noise part of the local field as the sum of two correlated
terms to which one can apply the CLT but one keeps the Gaussian assumption
throwing away possible non-Gaussian noise. For more details we refer to
\cite{O95}.
This further refined theory gives a better explanation of the dynamics
and the basin of attraction. Moreover, the storage capacity resulting
from this method is in good agreement with the results from computer
simulations.

However, we recall that the a priori Gaussian approximation is only
valid when  retrieval is successful. An improvement of the approximation
is obtained by taking into account explicitly the feedback: the network
state at time $t+1$, $\bsigma(t+1)$, depends on its previous states up to
$t-1$ namely  $\bsigma(0),\dots,\bsigma(t-1)$. One proposal has been to
assume two Gaussian peaks with variance calculated from the residual noise
and separated by an appropriately chosen distance \cite{HO90} but also here
the correlations between the network states
$\bsigma(t),\bsigma(t-1),\bsigma(t-2), \dots $ are only partially taken
into account.

More recently, one has studied the distribution of the local field without
any a priori assumption on the residual noise, trying to take into account
all correlations of the different terms of the sum
(\ref{eq:h:signal-noise:1}) using the insight gained before
\cite{PZ91,PZ91b} (see also \cite{B03} and references therein).

In the treatment of the correlations between the variables appearing in
this expression for the local field , one has to be very careful, because
even a small  dependence of order $\order{1/\sqrt{N}}$ may give rise to a
macroscopic  contribution after summation. Therefore, we rewrite the local
field
\begin{equation}
   \label{eq:h:splitting}
    h_i(t) = \hat h_i^\mu(t) + \frac{1}{\sqrt{N}} \xi_i^\mu r_N^\mu(t)
\end{equation}
where we have split of the part of $h_i(t)$ that depends strongly on
$\xi_i^\mu$, such that $\hat h_i^\mu$ only depends weakly on $\xi^\mu_i$.
Remark that the second term is small for large $N$,  so
\begin{eqnarray}
  \label{eq:s:deriv}
\sigma_i(t+1)
  & =& g(h_i(t)) \nonumber \\
  & = &g\left( \hat h_i^\mu(t) +
     \frac{1}{\sqrt{N}} \xi_i^\mu r_N^\mu(t) \right) \nonumber \\
  & = & g( \hat h_i^\mu(t))
      +  \left ( \frac{1}{\sqrt{N}}\xi_i^\mu r_N^\mu(t) \right )
         \left . \frac{\partial g}{\partial h_i} \right
	 |_{\hat h_i^{\mu}(t)}
      + \order{N^{-1}} \, .
\end{eqnarray}
Using this and the definition of the residual overlap in (\ref{eq:op:def}),
we get
\begin{equation}
    \label{eq:r:recur:N}
  r_N^\mu(t+1) \approx \tilde r_N^\mu(t) + \chi_N(t) \, r_N^\mu(t)
\end{equation}
where we have defined
\begin{equation}
    \label{eq:r:N}
   \tilde r_N^\mu(t)
      = \frac{1}{\sqrt N} \sum_i \xi_i^\mu g( \hat h_i^\mu(t)) \ ,
      \quad \mu >1 \,,
\qquad
    \chi_N(t)
    = \frac{1}{N} \sum_i
       \left . \frac{\partial g}{\partial h_i} \right
          |_{\hat h_i^{\mu}(t)} \ .
\end{equation}
Now, we take the limit $N\to\infty$. In this limit the density
distributions of $\hat h$ and $h$ are equal since they only differ
upto a factor $1/\sqrt{N}$.
Since $\hat h^\mu_i(t)$ only depends weakly on $\xi^\mu_i$, we assume
that in the limit  we can use the CLT for the first term. To the second
term in (\ref{eq:r:recur:N}), we  apply the LLN arriving at
\begin{equation}
  \label{eq:r:recur}
      r^\mu(t+1) = \tilde r^\mu(t) + \chi(t) \, r^\mu(t)
\end{equation}
with $\tilde r^\mu(t)$ a normally distributed random variable
$\distnorm(0,1)$ and
\begin{equation}
   \label{eq:chi:def}
\chi(t)
  = \avgg{ 
  \Bigg\langle\left . \frac{\partial g}{\partial h} \right |_{h(t)}
  \Bigg\rangle_{\boldsymbol{h}}
      }
  = \avgg{
  \int \prod_{s=0}^t d{h(s)} \, P(h(0), h(1), \ldots, h(t) )
      \left . \frac{\partial g}{\partial h} \right |_{h(t)}
     }
\end{equation}
where the average denoted by $\langle.\rangle_{{\boldsymbol h}}$ is over the
distribution of the local field with probability
density  $P(h(0), h(1), \ldots, h(t))\equiv  P({\boldsymbol{h}})$ and
where $\left \langle \! \left \langle \cdot \right \rangle \! \right
\rangle$ denotes the average over the initial conditions and the
condensed pattern.

Next, in order to find the distribution of the local field in the
thermodynamic limit, we start from (\ref{eq:h:signal-noise}) and use
(\ref{eq:op:def}) and (\ref{eq:r:recur:N})
\begin{eqnarray}
h_i(t+1)
  & =& \xi^\mu_i m_N(t+1)
     + \frac{1}{N} \sum_{\mu\ne1} g(\hat h_i^\mu(t))
     + \frac{1}{N} \sum_{\mu\ne1} \sum_{j \ne i}
        \xi_i^\mu \xi_j^\mu g(\hat h_j^\mu(t)) \nonumber \\
  & + & \chi_N(t) \frac{1}{\sqrt N} \sum_{\mu\ne1} \xi_i^\mu r^\mu_N(t)
     - \alpha \sigma_i(t+1) \, .
\end{eqnarray}
The first term on the r.h.s., using (\ref{eq:blabla}) is clear. In the
second term, we can replace
$\hat h^\mu_i(t)$ by $h^\mu_i(t)$ leading to a contribution
$\alpha \sigma_i(t+1)$.  The third term is a sum of independent random
variables and by the CLT it converges to $\distnorm(0, \alpha)$. In the
fourth term, we employ (\ref{eq:h:signal-noise:1}).
From this we obtain, omitting the site index $i$
\begin{equation}
  \label{eq:h:recur}
   h(t+1)= \xi^\mu m(t+1)
        + \chi(t)
           \left [
                h(t) - \xi^\mu m(t) + \alpha \sigma(t)
                                                    \right]
        + \distnorm(0,\alpha) \,.
\end{equation}
In this way  we arrive at the recursion relation (\ref{eq:h:recur}) for
the local field and (\ref{eq:r:recur}) for the residual overlap.
We still want a recursion relation for the overlap
by using the dynamics (\ref{eq:dyn:det:gen}) starting from
(\ref{eq:op:def})
 \begin{equation}
      m(t) = \avg{ 
        \av{ \xi^1 \, g(h(t-1))  }_{{\boldsymbol{h}}}
	 } \,.
\end{equation}

Finally, from (\ref{eq:h:recur}) it is clear that the local field
consist out of a discrete part and a normally distributed part
\begin{equation}
  \label{eq:M:def}
  h(t+1) = M(t) + \distnorm(0,\alpha D(t))
\end{equation}
where $M(t)$ can be found by iterating the recursion relation for the local
field, i.e.
\begin{equation}
   M(t) = \xi^1 m(t) + \alpha \sum_{s=0}^{t-1}
        \left( \prod_{s'=s}^{t-1} \chi(s') \right) \sigma(s) \ .
\end{equation}
The variance of the noise in (\ref{eq:M:def}) can be calculated by
using (\ref{eq:r:recur})
\begin{equation}
\label{eq:V:recur}
\mbox{Var}(r^\mu(t+1))
  = D(t+1)
  = 1 + \chi^2(t) \, D(t) +
       2 \chi(t) \, \mbox{Cov}(\tilde r^\mu(t), r^\mu(t)) \, .
\end{equation}

We still have to write out the  probability density of
the local field used to define $\chi(t)$ in (\ref{eq:chi:def}).  The
evolution equation tells us that $\sigma(t)$ can be replaced by
$g(h(t-1))$, such that the second term of $M(t)$ in (\ref{eq:M:def}) is
the sum of step functions of correlated variables.  These variables are
also correlated with the Gaussian part of the local field through the
dynamics.  Therefore, the local field can be seen as a transformation of
a set of correlated Gaussian variables $x(s)$ which we choose to normalize.
Defining the correlation matrix by $W(s,s')$, $s,s' \neq t-1$ we arrive at
the following expression for this  probability density
\begin{multline}
  \label{eq:h:sna}
  P({\boldsymbol{h}}) =
   \int \prod_{s=0}^{t-2}d x(s)  dx(t)
     \,    \delta\left(h(t) - M(t) - \sqrt{\alpha D(t)} x(t) \right) 
\\
     \times 
        \frac{1}{\sqrt{\det(2 \pi {\boldsymbol{W}})}}
        \exp\left(-\frac{1}{2}
        \sum_{s, s'=0 \atop s,s' \neq t-1}^t
                    x(s) {\boldsymbol{W}}^{-1}(s,s') x(s') \right) \, .
\end{multline}

This concludes the SNA treatment of the Little-Hopfield model at zero
temperature.
The above equations form a recursive scheme in order to calculate the
dynamical  properties of the system up to an arbitrary time step. The
practical difficulty which remains, certainly after a few time steps
is the explicit calculation of the correlations.

It is possible to extend the method to arbitrary temperatures by
introducing auxiliary thermal fields to express the stochastic dynamics
within the gain function formulation of the deterministic dynamics
\cite{PZ90}
\begin{equation}
  \label{eq:dyn:det}
    \sigma(t+1) = g(h(t) + T \gamma(t))
\end{equation}
where the $\gamma(t)$ are independent and identically distributed
with probability density
\begin{equation}
\label{distribgamma}
   P(\gamma(t)) = \frac{1}{2} \left(1 - \tanh^2( \gamma(t)) \right)\, .
\end{equation}
One then averages the zero temperature results over the auxiliary field
$\gamma(t)$. This alters the equations in a non-trivial
way but such that the idea of the derivation can be completely retained.
At this point we remark that in the derivation of the local field
distribution one has to replace $\sigma(t)$ by $g(h(t-1))$ with $h(t-1)$
given by (\ref{eq:M:def}). For arbitrary temperatures one has to replace
 $\sigma(t)$ by
$g(h(t-1) + T\gamma(t-1))$ and the average over the $\gamma(t)$ then
enters at the same level as the average over the noise.
For more details we refer to the literature \cite{PZ90}.

\subsection{Explicit results for the first four time steps}

In order to compare with the results of the GFA method to be explained in
the following section it is useful to recall the SNA results for the first
few time steps as found in \cite{BJS98} (and references therein). We only
write down explicitly the expressions for the overlap $m(t)$ at zero
temperature
\begin{eqnarray}
m(1) &=& \avg{\xi \int \gauss{z}\, g(\xi m(0) + \sqrt{\alpha D(0)} z)}
          \\
m(2) &=& \avg{\xi \int \gauss{z}\, g(\xi m(1) + \alpha \chi(0) \sigma(0)
           + \sqrt{\alpha D(1) z})}
       \\
m(3) &=& \avgl{ \xi \int \gauss{W^{0,2}}{(x,y)} \,
             g(\xi m(2) + \alpha \chi(1)[
              g(\xi m(0) } \nonumber \\
   && \qquad  \avgr{ + \sqrt{\alpha D(0)} x) + \chi(0) \sigma(0)
        ] + \sqrt{\alpha D(2)} y)
   }\\
m(4) & = &  \avgl{
   \xi \int \gauss{W^{0,1,3}}{(x,y,z)} \, g\Big[ \xi m(3)
   }
          \nonumber  \\
    &&\qquad  + \alpha \chi(2) \,
         g( \xi m(1) + \alpha \chi(0) \sigma(0)
                               + \sqrt{\alpha D(1)}y
                                                   )
                \nonumber \\
   &&\qquad  + \alpha \chi(2)\chi(1) \,
           g( \xi m(0) + \sqrt{\alpha D(0)}x
                                                   )
                       \nonumber  \\
    &&\qquad 
          + \alpha \chi(2)\chi(1)\chi(0) \sigma(0)  \nonumber
	  \\
    &&\qquad  \avgr{+ \sqrt{\alpha D(3)} z
                                        \Big] 
		}
					 \, .
    \label{eq:m(4):SNA}
\end{eqnarray}
Here $\gauss{z}$ is the Gaussian measure with variable $z$
while $\gauss{W}{(x_1,\dots,x_t)}$ is the multidimensional Gaussian measure
with correlation matrix ${\boldsymbol W}$ as it appears in (\ref{eq:h:sna}).
For the explicit form of the expressions for the variance of the
residual overlap, the function $\chi$ and the correlations we refer to
the literature \cite{BJS98}.

\section{Generating functional approach to the Little-Hopfield model}

\subsection{The effective local field}

The idea of the GFA aproach to study dynamics \cite{C01,MSR73,
Dom78}
is to look at the probability to find a certain microscopic path in time.
The basic tool to study the statistics of these paths is the generating
functional
\begin{equation}
Z[{\bpsi}] = \sum_{\bsigma(0), \dots, {\bsigma}(t)}
             P[{\bsigma}(0), \dots, {\bsigma}(t)]
    \, \prod_{i=1}^N \prod_{s=0}^t e^{-i \, \psi_i(s) \, \sigma_i(s) }
\end{equation}
with $P[{\bsigma}(0), \dots, {\bsigma}(t)]$ the probability
to have a certain path in phase space
\begin{equation}
P[{\bsigma}(0), \dots, {\bsigma}(t)]
 = P[{\bsigma(0)}] \prod_{s=0}^{t-1} W[{\bsigma}(s+1)|{\bsigma}(s)]
\end{equation}
and $W[{\bsigma}|{\bsigma}']$ the transition probabilities from
$\bsigma'$ to  $\bsigma$.  Looking back at (\ref{eq:dyn:glauber}) and
assuming parallel dynamics we have
\begin{equation}
W[{\bsigma}(s+1)|{\bsigma}(s)]
  = \left .
      \prod_{i} \frac{e^{\beta \sigma_i(s+1) h_i(s)}}
                       {2\cosh(\beta h_i(s))}
    \right|_{h_i(s) = \sum_j J_{ij} \sigma_j(s) + \theta_i(s)}
    \label{transprob}
\end{equation}
where we have introduced a time-dependent external field $\theta_i(s)$ in
order to  define a response function.

One can find all the relevant order parameters, i.e., the overlap
$m(t)$, the correlation function $C(t,t')$ and the response function
$G(t,t')$, by calculating appropriate derivatives of the above functional
and letting ${\bpsi}=\{\psi_i\}$ tend to zero afterwards
\begin{eqnarray}
m(t)    & =& i \lim_{\bpsi\to0}
           \frac{1}{N} \sum_{i}\xi_i
           \frac{\delta Z[\bpsi]}{\delta\psi_i(t)} \\
G(t,t') & =& i \lim_{\bpsi\to0}
           \frac{1}{N} \sum_{i}
           \frac{\delta^2 Z[\bpsi]}{\delta\psi_i(t) \delta\theta_i(t')} \\
C(t,t') & =& - \lim_{\bpsi\to0}
           \frac{1}{N} \sum_{i}
           \frac{\delta^2 Z[\bpsi]}{\delta\psi_i(t) \delta\psi_i(t')}\,
	   .
\end{eqnarray}

In the thermodynamic limit one expects the physics of the problem to be
independent of the quenched disorder and, therefore, one is
interested in derivatives of $\overline{Z[{\bpsi}]}$ with the
overline denoting the average over this disorder, i.e., all pattern
realizations.
This results in an effective single spin local field given by
\begin{equation}
   \label{eq:h}
    h(t) = \xi\,m(t) + \alpha \sum_{s=0}^{t-1} R(t,s) \, \sigma(s)
         + \sqrt{\alpha} \nn(t)
\end{equation}
with $\eta(t)$ temporally correlated noise with zero mean and correlation
matrix
\begin{equation}
  \bD = (\one-\bG)^{-1} \bC (\one - \bG^\dagger)^{-1}
\end{equation}
and the retarded self-interaction
\begin{equation}
   \bR = (\one-\bG)^{-1}\bG \, .
\end{equation}
We refer to \cite{C01} for further details.
The order parameters defined above can be written as
\begin{eqnarray}
m(t)    & = & \avs{ \av{ \xi \sigma(t) }_\star } \\
C(s,s') & = & \avs{ \av{ \sigma(s) \, \sigma(s') }_\star} \\
G(s,s') & = & \avgg{\av{\frac{\partial \sigma(s)}{\partial \theta(s')}
                                     }_\star} \, .
 \label{eq:ordergfa}
\end{eqnarray}
The average over the effective path measure is given by
\begin{equation}
    \label{eq:noise:dist}
\av{f}_\star
=
  \mbox{Tr}_{\{\sigma(1),\ldots, \sigma(t) \}}
    \int d { \bfeta}
    \, P({\bfeta})
    \, P({\bsigma} | {\bfeta})
    \, f \ ,
  \end{equation}
where $d { \bfeta}= \prod_s \eta(s)$ and with
\begin{eqnarray}
    \label{eq:s|eta}
P({\bsigma}|{\bfeta})
 =
   \prod_{s=0}^{t-1}
    \left .
      \frac{\exp(\beta \sigma(s+1) h(s))}{2 \cosh(\beta h(s)) }
    \right |_{h(s) = \xi\,m(s) + \sum_{p} R(s,p) \, \sigma(p) + \nn(s)}
   \\
P({\bfeta})
=
     \frac{1}{\sqrt{\det(2\pi\bD)} }
 \exp
   \left (
     -\frac{1}{2} \sum_{s,s'=0}^{t-1} \eta(s) \, \bD^{-1}(s,s') \, \eta(s')
     \right ) \ .
     \label{eq:eta}
\end{eqnarray}
The average denoted by the double brackets is again (as in the SNA) over
the condensed pattern and initial conditions.
We remark that $G(s,s') = 0$ for $s\le s'$ and
$D(s,s')=D(s',s)$, and that for all $s<t$
\begin{equation}
    \label{eq:G:T>0}
G(t,s) = \beta \left\langle\!\left\langle
     \av{\sigma(t)[\sigma(s+1) - \tanh(h(s))]}_\star
                 \right\rangle\!\right\rangle
\end{equation}
where $h(s)$ is given by (\ref{eq:h}).

\subsection{An alternative stochastic description}

The GFA at arbitrary temperatures starts from the transition probabilities
(\ref{transprob}) following from the Glauber dynamics
(\ref{eq:dyn:glauber}).
In order to make the comparison with the SNA later on we want to introduce
an  alternative description starting from the deterministic dynamics as in
(\ref{eq:dyn:det})-(\ref{distribgamma}).

Starting from (\ref{eq:s|eta}) and (\ref{eq:eta}), taking the limit of
zero temperature and introducing the auxiliary thermal field $\gamma(t)$
both formulations are equivalent. Indeed, suppose that we want to calculate
the effective path average of a
general function $f$  using the alternative description.  Writing out the
expression for this average  (the average over the
condensed pattern and initial conditions is not relevant here)
\begin{eqnarray}
     \label{eq:av:alt}
&& \hspace{-1cm}\int d \bgamma  P(\bgamma) \left (
  \lim_{\beta \to \infty} \av{f({\bsigma})}_\star
        \right)_{h \to h + T \gamma} 
  \nonumber
  \\ &&=
  \mbox{Tr}_{\{\sigma(1), \ldots, \sigma(t)\}}
    \int d {\bgamma}
    \int d{\boldsymbol{h}}
    \int d {\bfeta}
    \,  \, P({\bfeta}) \nonumber \\
&& \qquad  \prod_{s=0}^{t-1}
       \left[\delta\left(
               h(s) - \xi m(s) - \alpha \sum_{s'} R(s,s') \sigma(s')
             - \sqrt{\alpha}\eta(s) - T \gamma(s)
         \right) \nonumber \right. \\
&& \qquad  \left. \frac12 \left( 1-\tanh^2\left({\gamma(s)}\right) \right)
       \Theta(\sigma(s+1) \, h(s)) \, \right]
       f({\bsigma})
\end{eqnarray}
where $\gamma$ is the thermal field. Since the latter only occurs inside the
$\delta$-function and the  probability density, we can evaluate
the integral over $\gamma$.  Then the integration over the local field becomes
\begin{multline}
\prod_{s=0}^{t-1}
\int d h(s) \,
  \Theta(\sigma(s+1) \, h(s)) \\ 
  \left [ 1-\tanh^2 \left(
                   h(s) - \xi m(s) - \alpha \sum_{s'} R(s,s') \sigma(s')
               - \sqrt{\alpha}\eta(s)
                    \right) \right] \, .
\end{multline}
This integral can be evaluated, and yields
\begin{equation}
   \prod_{s=0}^{t-1}
    \left .
      \frac{\exp(\beta \sigma(s+1) h(s))}{2 \cosh(\beta h(s)) }
    \right |_{h(s) = \xi\,m(s) + \alpha \sum_{s'} R(s,s') \, \sigma(s')
                         + \sqrt{\alpha}\nn(t)}
\end{equation}
which is exactly the same  as (\ref{eq:s|eta}) showing that we can use
both representations for the dynamics.

Using this alternative description for the thermal dynamics it is
straightforward to write the distribution of the local field as
\begin{eqnarray}
   \label{eq:h:alt:t}
P({\boldsymbol{h}}) &=&  
    \int d {\bgamma} d {\bfeta}
    \, P({\bfeta}) \, P({\bgamma}) \nonumber \\
 &&   \left .
  \delta(h(t) - \xi\,m(t) - \alpha \sum_{s=0}^{t-1} R(t,s) \, \sigma(s)
        - \sqrt{\alpha}\nn(t) - T \gamma(t))
    \right |_{\sigma,h}
\end{eqnarray}
where we  denote by $|_{\sigma,h}$ the substitutions $\sigma(s) = g(h(s-1))$
and $h(s)=\xi\,m(s) + \sum_{s'} R(s,s') \, \sigma(s') +
\sqrt{\alpha}\nn(s) + T \gamma(s)$  for all $s \leq t$.
The above discussion  makes it trivial to perform the zero temperature
limit in the GFA and  also shows that it is enough to look at the
relation between both methods at zero temperature in order to be able to
compare them in the whole phase diagram, because the way to extend
the methods to finite temperature is completely equivalent.

\subsection{Recursive scheme}

The aim of this subsection is to derive recursion relations for the local
field and also for the noise appearing in the local field starting from
the GFA. In this way we want to
gain more insight into the equations at hand and make a detailed comparison
between the SNA and GFA.

From expression (\ref{eq:h}) for the effective local field, it is not
immediately obvious how $h(t)$ depends on $h(s)$, $s < t$. First, we
want an expression for the retarded self-interaction $\bR$ as a function of
the previous time step(s). To this end we write the response matrix at
time $(t+1)$ in the following way
\begin{equation}
      \label{eq:G:decomp}
\bG_{t+1} =
  \left (
     \begin{array}{cc}
     \bG_t & 0 \\
     \bg_{t+1} & 0
     \end{array}
  \right )
\end{equation}
where $\bg_{t+1}$ is the following vector of dimension $t$
\begin{equation}
    \bg_{t+1} = (G(t+1,0), G(t+1,1), \dots, G(t+1,t))
\end{equation}
and $\bG_t$ is the response matrix at time $t$.  It is clear that adding a
time step adds a row and a column to the matrix while it leaves the other
elements unaltered. From the decomposition of the response matrix, we
calculate $\bR_{t+1}$ (see, e.g.,  \cite{HJ85})
\begin{equation}
      \label{eq:R:recur}
\bR_{t+1} =
  \left (
    \begin{array}{cc}
      \bR_t & 0 \\
      \bg_{t+1}(\one + \bR_{t}) & 0
    \end{array}
  \right )
\end{equation}
and use this expression in (\ref{eq:h})  to arrive at
\begin{equation}
      \label{eq:h:recursion}
\nonumber h(t) =
  \xi \, m(t)
  + \sum_{s=0}^{t-1} G(t,s)
      \left[
        h(s) - \xi \, m(s) + \alpha \sigma(s)
      \right ]
  + \sqrt{\alpha} \sum_{s=0}^{t} (\one - \bG)(t,s) \, \nn(s) \, .
\end{equation}
This is one of the main results of this Section and will be used as the
basis of comparison between the GFA and SNA approaches. It
leads naturally to the definition of a modified noise variable $\tn$
\begin{equation}
     \label{eq:etat}
\tn(t) = \sum_{s=0}^{t} (\one - \bG)(t,s) \, \nn(s)\, ,
\qquad
\av{\tn(s) \tn(s')} = \bC(s,s')
\end{equation}
and it follows that the covariance matrix of the noises, $\bB$, is
given by
\begin{equation}
 B(s,s') = \av{\nn(s) \, \tn(s')}
= \sum_{s''}^{t} (\one-\bG^\dagger)(s'',s') \av{\nn(s)\,\nn(s'')}
= [(\one - \bG)^{-1} \bC](s,s')\, .
\end{equation}
The correlation matrix of this transformed noise variable is clearly
simpler than the original one and is in fact the correlation matrix of
the spins.

Next, we derive recursion relations for the noise in a similar way.
Analogous to  (\ref{eq:G:decomp}) we write
\begin{equation}
          \label{eq:C:decomp}
\bC_{t+1} =
  \left (
    \begin{array}{cc}
      \bC_t &  \bc_{t+1}^\dagger \\
      \bc_{t+1} & 1
    \end{array}
  \right )
\end{equation}
where $\bc_{t+1}$ is the following vector of dimension $t$
\begin{equation}
    \bc_{t+1} = (C(t+1,0), C(t+1,1), \dots, C(t+1,t))\, .
\end{equation}
One then finds
\begin{equation}
          \label{eq:D:recursion}
 \bD_{t+1}
  = \left (
    \begin{array}{cc}
      \bD_t
        &
      \bD_t \bg^\dagger_t + (\one - \bG_t)^{-1} \bc^\dagger_{t+1}
        \\ \\
      \bg_t \bD_t + \bc_{t+1} (\one - \bG^\dagger_t)^{-1}
        &
      1 + \bg_t \bD_t \bg^\dagger_t +
             2 \bc_{t+1} (\one - \bG_t)^{-1} \bg^\dagger_t
    \end{array}
    \right )
\end{equation}
Again, going one time step further implies adding one row and column to the
matrix while the rest of the matrix remains unchanged.
From (\ref{eq:D:recursion}) we find a relation for the variance of the
noise at time $t+1$
\begin{eqnarray}
         \label{eq:eta:recur}
D(t+1,t+1)
& = & 1 + \sum_{s,s'=0}^{t} G(t+1,s) \, D(s,s') \, G(t+1, s')
    \nonumber \\
     &&\qquad \qquad  + 2 \sum_{s=0}^{t} G(t+1,s) \, B(s,t+1)
\end{eqnarray}
as a function of the variance of the noise at time $s<t$ and other
quantities. Apparently, the right-hand side of this equation does still
depend on $t+1$ and, therefore, does not seem to be really a recursion
relation.  However, as we will show, the quantities on the right-hand side
can be calculated without having full information about time $t+1$,
e.g., $G(t+1,s)$ can be obtained without knowing $D(t+1,t+1)$.

\section{SNA versus GFA}

\subsection{Short-memory approximation}

Comparing the expressions for the effective local field
(\ref{eq:h:recursion}) and (\ref{eq:h:recur}) we notice that the first
one contains a sum over all time steps up to $t-1$ while the second one
only contains $t-1$ itself. Therefore, we introduce the following
approximation
\begin{equation}
    \label{eq:ma}
\sum_{s=0}^{t-1} G(t,s) \, h(s)
\approx
    \ma{X}(t-1) \, \ma{h}(t-1)
\equiv
    \sum_{s=0}^{t-1} \ma{G}(t,s) \, \ma{h}(s)
\end{equation}
where we have defined $\ma{X}$, $\ma{h}$ and $\ma{\bG}$ in this way.
The approximated matrix $\ma{\bG}$ now has a simple form
\begin{equation}
\ma{\bG} =
  \left (
  \begin{array}{ccccc}
    0      &  0      &  \dots   &   0     &   0  \\
    \ma{X}(0)   &  0      &  \dots   &   0     &   0  \\
    0      &  \ma{X}(1)   &  \dots   &   0     &   0  \\
    \vdots &  \vdots &  \vdots  &   0     &   0  \\
    0      &  0      &  0       &  \ma{X}(t-1) &   0  \\
  \end{array}
  \right )
\end{equation}
This approximation reduces the expression for the local field
(\ref{eq:h:recursion}) to
\begin{equation} \label{eq:h:ma}
\ma{h}(t+1) =
  \xi \, \ma{m}(t+1)
  + \ma{X}(t)
      \left[
        \ma{h}(t) - \xi \, \ma{m}(t) + \alpha \ma{\sigma}(t)
      \right ]
  + \sqrt{\alpha} \ma{\tn}(t+1) \, .
\end{equation}
Furthermore, the modified noise equation (\ref{eq:etat}) simplifies to
\begin{equation} \label{eq:etat:ma}
\ma{\tn}(t) = \ma{\nn}(t) - \ma{X}(t-1) \ma{\nn}(t-1)
\end{equation}
and the variance of the noise itself (\ref{eq:eta:recur}) becomes
\begin{equation} \label{eq:eta:rec:ma}
\ma{D}(t+1,t+1)
 = 1 + \ma{X}^2(t) \, \ma{D}(t,t) + 2 \ma{X}(t) \, \ma{B}(t,t+1)\, .
\end{equation}
Another consequence of this approximation is that the discrete part in
the local field (\ref{eq:M:def}) can be written down  or, in
other words, the retarded self-interaction $\bR$ can be calculated
explicitly to be
\begin{equation}
         \label{eq:M:approx}
R(t,s)
   =   \prod_{s'=s}^{t-1} \chi(s') \, .
\end{equation}

Comparing the recursion relations (\ref{eq:h:recur}) and (\ref{eq:h:ma})
for the local field, those for the residual overlap,
(\ref{eq:r:recur}) and (\ref{eq:etat:ma}), and  the ones for the variance
of the residual overlaps, (\ref{eq:V:recur}) and (\ref{eq:eta:rec:ma}), we
find that they are formally the same by taking
\begin{equation}
\ma{X}(t) = \chi(t)\, ,
\qquad
\ma{\tn}(t) = \tilde r^\mu(t-1)\, ,
\qquad
\ma{\nn}(t) = r^\mu(t) \, .
\label{eq:bla}
\end{equation}

Next, we would like to know  what is the physics  behind this
approximation. In fact, all solutions to (\ref{eq:ma}) can be called
k-th order short-memory approximations because they approximate the
feedback
\begin{equation}
\label{eq:smak}
  \sum_{{s=0}}^{t-1} G(t,s) \, h(s) \sim
    \left ( \sum_{s=t-k \atop s \geq 0}^{t-1}
                                  G(t,s) \right ) \, h(t-1)\, .
\end{equation}
These approximations take into account that responses, in general, decrease
very fast as a function of time. As we will show, the first order
approximation obtained for $k=1$ and simply called the short-memory
approximation, corresponds to the SNA
\begin{equation}
  \sum_{s=0}^{t-1} G(t,s) \, h(s)  \sim
    G(t,t-1) \, h(t-1)\,,
\qquad  \chi(t)=G(t+1,t) \,.
\end{equation}
We remark that these approximations all have a different $\chi(t)$ so
that we can use the latter in order to distiguish between them. We calculate
the $\chi(t)$ corresponding to $k=1$, starting from the alternative
stochastic description of the GFA (forgetting about the average over the
initial conditions and condensed pattern). We find
\begin{eqnarray}
 G(t+1,t)
  &=& 
    \int d {\bfeta}\,  d {\bgamma}
     \, P({\bfeta}) \, P({\bgamma})
      \left .
          \frac{\partial \sigma(t+1)}{\partial h(t)}
       \right|_{\sigma,h} \nonumber \\
  &=&  \int d {\bfeta} \,  d {\bgamma}
     \, P({\bfeta}) \, P({\bgamma})
      \left .
          \frac{\partial g(h(t))}{\partial h(t)}
       \right|_{\sigma,h} \nonumber \\
  &=&  \left .
       \int d {\bfeta} \,  d {\bgamma}
       \, P({\bfeta}) \, P({\bgamma})
         \,   \delta(h(t))
       \right|_{\sigma,h}\nonumber \\
  &=&  
       \int \left( \prod_{s=0}^{t} \frac{d \eta(s) d \gamma(s)}
             {\det(2\pi \bD)} \right) \nonumber\\
  && \qquad
      \left . 
      \exp\left(
                 -\frac{1}{2} \sum_{s,s'}^{t} \eta(s)
		      \bD^{-1}(s,s') \eta(s')
              \right)
     \, P({\bgamma})
         \, \delta(h(t))
       \right|_{\sigma,h}\, . \nonumber
\end{eqnarray}
Recalling the distribution of the local field at time $t$
(\ref{eq:h:alt:t}), and taking the temperature to be zero (which means
vanishing $\gamma$), this expression  already resembles the analogous
expression (\ref{eq:h:sna}) in the SNA approach.
Furthermore, one sees that $\eta(t-1)$ does not occur in the integrand
of this expression. As a consequence, one can show that the matrix $\bD$
can be replaced by
\begin{equation}
\overline{D}(s,s') = \avv{\eta(s) \eta(s')}{\eta}\, ,
          \quad
           s,s' \ne t-1
\end{equation}
or in more detail
\begin{equation}
\overline{\bD} =
  \left (
  \begin{array}{ccccc}
    D(0,0)    &  D(0,1)   &  \dots   &   D(0,t-2)    &  D(0,t)  \\
    D(1,0)    &  D(1,1)   &  \dots   &   D(1,t-2)    &  D(1,t)  \\
    \vdots    &  \vdots   &  \vdots  &   \vdots      &  \vdots  \\
    D(t-2,0)  &  D(t-2,1) &  \dots   &   D(t-2,t-2)  &  D(t-2,t)  \\
    D(t,0)    &  D(t,1)   &  \dots   &   D(t,t-2)    &  D(t,t)   \\
  \end{array}
  \right )
\end{equation}
One can then verify that ${G(t+1,t)}$, for zero temperature, is then
exactly the same
as the expression for $\chi(t)$ in the SNA approach, viz.
\begin{multline}
 G(t+1,t)
  = 
       \int \left( \prod_{s=0 \atop s \ne t-1}^{t}
         \frac{d \eta(s)}{\det(2\pi \overline{\bD})} \right) \\
       \, \left .
         \exp\left(
                 -\frac{1}{2} \sum_{s,s' \atop s,s' \ne t-1}^{t} \eta(s)
               \overline{\bD}^{-1}(s,s') \eta(s')
              \right)
         \, \delta(h(t))
       \right|_{\sigma,h}  .
\end{multline}
In conclusion, compared with the exact GFA method we find that the SNA
approach is a short-memory approximation, in which the response from
earlier times is not taken into account.

\subsection{Discussion}

The origin of the approximation inherent in the SNA approach as
discussed above lies in the treatment of the residual noise
$\tilde r_N^\mu(t)$ (recall (\ref{eq:r:N})).
We have assumed that after taking out the term
${N}^{-1/2} \xi_i^\mu r_N^\mu(t)$  of the local field,
$\xi_i^\mu$ and $\hat h_i^\mu(t)$ are only
weakly correlated, and therefore we have applied the CLT to find that
$\tilde r_N^\mu(t)$ converges to a normal distribution.  Comparing with
the GFA it
appears that we took out only part of the correlations between
$\sigma_i(t)$ and $\xi_i^\mu$, those coming from the previous time step.
In a fully connected system however, there are not only
feedback loops of length 1 and 2, but of arbitrary length, although
they may be less probable \cite{BKS90}.

Therefore, one would expect the SNA method to approximate the dynamics
already from the second time step onwards. However, we can show
that $\bG(t,t-2)$ is zero such that the second time step is still exact.
Moreover, some of the order parameters in the third time step, e.g., the
overlap only involve noises up to the second time step, so that they
are also correct for the third time step. 

In order to show that $G(t,t-2)=0$ we proceed as follows
(recall (\ref{eq:G:T>0}))
\begin{eqnarray}
           \label{eq:G:t-2}
G(t,t-2)
& = &  \frac{\partial \avv{\sigma(t)}{\star}}{\partial \theta(t-2)}
  \nonumber \\
& = & \beta \avv{\sigma(t)
       \left[ \sigma(t-1) - \tanh(\beta h(t-2)\right)]}{\star}\, .
\end{eqnarray}
Considering the effective path average  the sum over $\sigma(t)$ can
be done  explicitly
\begin{equation}
\sum_{\sigma(t)=\pm 1}
   \sigma(t)
   \frac{e^{\beta \sigma(t) h(t-1)}}{2 \cosh(\beta h(t-1)}
 = \tanh(\beta h(t-1)) \, .
\end{equation}
Moreover, the expression for the distribution only contains summations
over $\sigma(0)$ up to $\sigma(t-1)$ and there is only one term
left that contains $\sigma(t-1)$ so
\begin{equation}
\sum_{\sigma(t-1)=\pm 1}
   \sigma(t-1)
   \frac{e^{\beta \sigma(t-1) h(t-2)}}{2 \cosh(\beta h(t-2))}
 = \tanh(\beta h(t-2)) \, .
\end{equation}
Remark that further sums over spins can not be done since $h(t-1)$
contains the spins at times $s \leq t-2$.
This shows that $G(t,t-2) = 0$.

\subsection{Accuracy in retrieval}

First, for $\alpha=0$, one can easily verify that the GFA analysis yields
\begin{equation}
      \label{eq:G:alpha=0}
\bG =
  \left (
  \begin{array}{ccccc}
    0      &  0      &  \dots   &   0     &   0  \\
    \chi(0)   &  0      &  \dots   &   0     &   0  \\
    0      &  \chi(1)   &  \dots   &   0     &   0  \\
    \vdots &  \vdots &  \vdots  &   0     &   0  \\
    0      &  0      &  0       &  \chi(t-1) &   0  \\
  \end{array}
  \right )
\end{equation}
implying that the short-memory approximation, i.e., the SNA  is exact.
The reason is  that the local field at time $t$ does not depend on previous
times because the terms in the retarted self-intereaction and the
noise are proportional to $\alpha$. The only correlation that remains in
the system comes from  the dependence of the spins on the local field at
the previous time step.
Consequently, we expect the SNA to be a very good approximation to
the exact dynamics for small loading capacities. Some extra reasons to do
so is that for small loading capacities, the convergence time to the
attractor is very small, only a few time steps, and the SNA is exact up to
time step $3$. Moreover, to write down the evolution equations for the
order parameters like the overlap one performs averages over the noise.
Therefore, not surprisingly, the SNA results for the first few time steps
coincide with numerical simulations as has been reported in the literature
before. Only discrepancies of the order ${\cal O}(10^{-3})$ that are of the
same magnitude as the finite size effects in the simulations, and hence
not conclusive,  have been  observed (see, e.g., \cite{B03} and references
therein).

Analytically, of course, one notices the difference from the fourth time
step onwards.  Starting from the exact  GFA approach and writing the
expression in a convenient form for comparison, we get at zero
temperature
\begin{eqnarray}
        \label{eq:m(4):GFA}
m(4) & = &
  \avgl{
    \xi \int \gauss{W^{0,1,3}}{(x,y,z)} \, g\Big[ \xi m(3) 
  }
          \nonumber   \\
    &&\qquad  \left. \left.   + \alpha \tilde \chi(2) \,
         g( \xi m(1) + \alpha \chi(0) \sigma(0)
                               + \sqrt{\alpha D(1)}y
                                                   )
                \nonumber \right. \right.\\
   &&\qquad \left. \left.   + \alpha \tilde \chi(2) \tilde \chi(1) \,
           g( \xi m(0) + \sqrt{\alpha D(0)}x
                                                   )
                       \nonumber \right. \right. \\
    &&\qquad\left. \left.
          + \alpha \tilde \chi(2) \tilde \chi(1) \tilde \chi(0) \sigma(0)
	  \underline{+ \alpha \bG(3,0) \sigma(0)}
	    \nonumber \right. \right.\\
    &&\qquad  \avgr{  + \sqrt{\alpha D(3)} z
                                        \Big]} \,
\end{eqnarray}
The only difference with the SNA result (\ref{eq:m(4):SNA}) is in the
term that is underlined.  It is present due to the fact that beyond
$t=3$, $\bR$ is not simply given by (\ref{eq:M:approx}).
We will show numerically in the next subsection that this difference is
small (e.g., of the order of 0.3\% upto 3\% for $T=0.1, \alpha=0.1$,
$m_0=0.4$ respectively $m_0=0.2$). It will be interesting to see how this
difference behaves for further time steps.

\subsection{Numerical results}

Now that we know precisely what the origin is of the approximation inherent
in the SNA method as usually applied, we can check numerically how
accurate it is for retrieval and also for spin-glass behaviour, although
from the point of view of neural networks one is not primarily interested
in the latter.

We first want to compare the limiting normal distribution of
$\tilde r_N^{\mu}$ in the SNA approach  with  simulations for different
time steps. This is done in Fig.~1 for two representative 
values of the capacity, i.e. $\alpha=.06$
which lies in the retrieval phase and $\alpha=.16$ which is above the critical
capacity and thus within the spinglass phase.

\begin{figure}[t]
\label{fig1}
\begin{center}
\noindent
\hfil
\includegraphics[width=.48\textwidth]{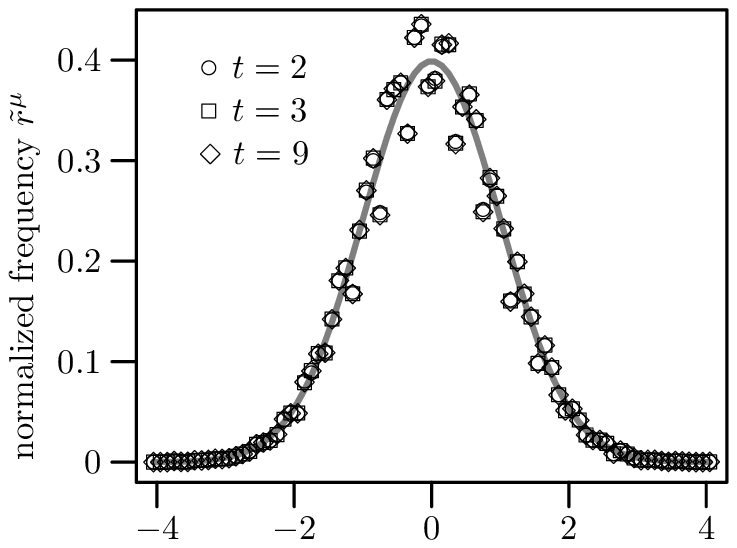}
\hfill
\includegraphics[width=.48\textwidth]{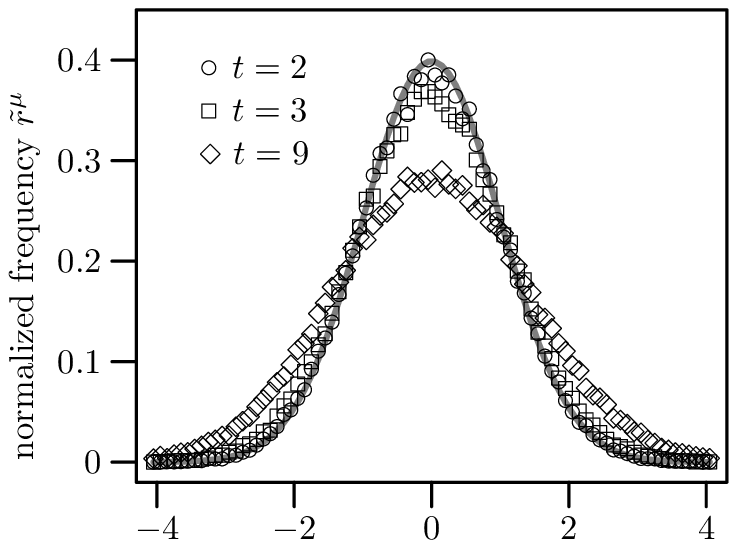}
\hfil
\\

\caption{Simulations for the distribution of the residual overlap, 
$\tilde r(t)$, for different time steps compared with
their theoretical distribution as calculated using the SNA (full curve). The
picture on the left is for $\alpha=0.06$ while the one on the right
is for $\alpha=0.16$ corresponding to
retrieval respectively spinglass behaviour.
Both pictures where made for $T=.1$, $m_0=.6$ using a finite size simulation
with $N=3000$ ($N=2000$) for the left (right) figure and averaging over 200 samples. } 
\end{center}
\end{figure}

We conclude that in the retrieval region ($\alpha_c < .135$ for $T=.1$) the
simulation results coincide quite good with the limiting SNA
distribution, while in the spin-glass region, the results for $t=3$  start to
divert systematically.  This is consistent with the results obtained in the
fully connected Blume-Emery-Griffiths neural network~\cite{BBSV03}.  Remark
that the distribution of $\tilde r^\mu$ starts to divert for $t=3$, while
$m(3)$ is still exact as discussed before.  

We also want to compare the evolution of the order parameters.
As a typical result we show in  Fig. 2 the overlap $m(t)$ as a function of
time using the method in
\cite{EO92}, precisely in order to avoid finite size effects.
\begin{figure}[t]
\begin{center}
\includegraphics{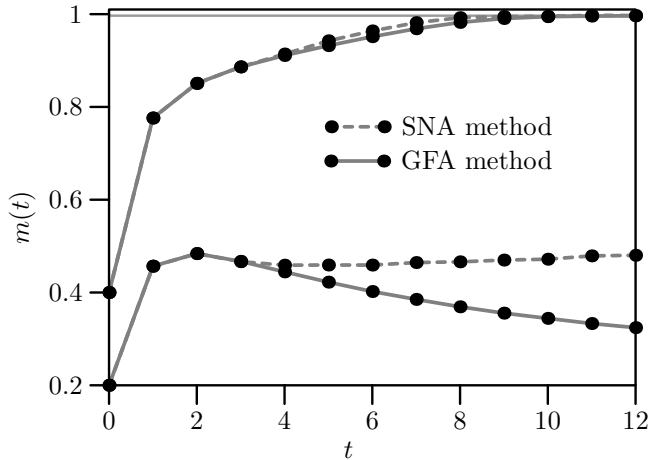}
\caption{The overlap order parameter as a function of time for the SNA
versus the GFA approach for $\alpha=0.1$, $T=0.1$ for retrieval behaviour
$m_0 =0.4$ and spin-glass behaviour $m_0 =0.2$. The thin line represents
the stationary limit for retrieval.}
\end{center}
\label{fig2}
\end{figure}
We see that the SNA results coincide with those of the GFA up to the third
time step, as shown analytically. For further time steps the results
are strongly dependent upon the parameters of the network determining
its behaviour: retrieval or spin-glass behaviour. The parameters chosen
in Fig.~2 are $\alpha=0.1$, $T=0.1$, and $m_0 =0.4$ respectively $m_0
=0.2$. For the first choice (the two upper curves) the system evolves to the
retrieval attractor and one observes that there is only a marginal difference
between the SNA and GFA method. This confirms the previous observations made
in the literature (\cite{B03} and references therein).  For the second
choice (the two lower curves) the initial overlap is too small such that we
 are outside the basin
of attraction for retrieval and, hence, we do not evolve to the attractor
(spin-glass behaviour). Here the SNA method, as used in the literature, does
not give good results for further time steps. We remark that we have
also compared these results with finite size simulations. These
simulations are not indicated on the figure because they lie almost
exactly on the GFA lines.

\section{The SNA revisited}

In this section we show how to treat the feedback correlations in the SNA
approach correctly.
Looking at the definition of the residual overlap at time $t$, i.e.,
\begin{equation} \label{eq:r:def:again}
r_N^\mu(t)
  = \frac{1}{\sqrt{N}} \sum_{i=1}^N \xi^\mu_i \sigma_i(t)
  \quad \mu > 1 \ ,
\end{equation}
we want to know how the dependence between spins and patterns 
evolves in the course of the dynamics.

In general, the local field at a certain time step $h_i(t)$ depends on
$\xi_i^\mu$ in two ways.  First, there is the correlation through the
occurrence of $\xi_i^\mu$ in the coupling matrix ($J_{ij}$).  We call this
type of correlations first order correlations because they are the most
obvious ones.  Second, there may be so-called second order correlations, being
correlations with pattern $\xi_i^\mu$ through the dynamics as a result of
feedback.  Second order correlations can only appear due to first order
correlations earlier in the dynamics.  Therefore, extracting the first order
correlations at every time step results in a system that does not depend on
$\xi_i^\mu$ anymore.

We define $\delta h_i^\mu(s)$ by
\begin{equation}
\delta h_i^\mu(s) 
   = -\frac{1}{N} \sum_j \xi_i^\mu \xi_j^\mu \sigma_j(s)
   = -\frac{1}{\sqrt{N}} \xi_i^\mu r^\mu(s) \ .
\label{eq:delta-h}
\end{equation}
In this way, adding $\delta h_i^\mu(t)$ removes the first order correlations
from the local field at time $t$, i.e.,
\begin{equation}
\tilde h_i^\mu(s) = h_i(s) + \delta h_i^\mu(s)
\end{equation}
such that $\tilde h_i^\mu(s)$ only depends on $\xi_i^\mu$ in second order.
Moreover, $\delta h_i^\mu(s)$ tends to zero as $1/\sqrt{N}$ in the 
thermodynamic limit, and can therefore be regarded as a perturbation to the
field $h_i(s)$.
We apply this perturbation for all time steps $s<t$ and find
\begin{equation}
\tilde \sigma^\mu_i(t) 
   =  \sigma_i(t) +
        \sum_{s=0}^{t-1} 
         \frac{\partial \sigma_i(t)}
              {\partial \tilde h_i^\mu(s)}\Bigg|_{\delta h_i^\mu(s)=0}
            \delta h_i^\mu(s) + \order{1/N} \ ,
\label{eq:s-recur}
\end{equation}
where $\tilde \sigma^\mu_i(t)$ is now completely independent of $\xi_i^\mu$.
We rewrite the derivatives by introducing an external field
$\theta_i(s)$,
\begin{equation}
\frac{\partial \sigma_i(t)}
     {\partial \delta h_i^\mu(s)}\Bigg|_{\delta h_i^\mu(s)=0}
= \frac{\partial \sigma_i(t)}
       {\partial \theta_i(s)}\Bigg|_{\theta_i(s)=0}
\label{eq:part-der} \ .
\end{equation}
Inserting (\ref{eq:delta-h}) in (\ref{eq:s-recur}) and using
(\ref{eq:part-der}) yields
\begin{equation}
\sigma_i(t)
   = \tilde \sigma^\mu_i(t)
       + \frac{\xi_i^\mu}{\sqrt{N}} 
          \sum_s 
            \frac{\partial \sigma_i(t)}
                 {\partial \theta_i(s)}\Bigg|_{\theta_i(s)=0}         
          r^\mu(s)+ \order{1/N} \, .
\end{equation}
We multiply both sides by $\xi_i^\mu/\sqrt{N}$ and sum over $i$, so that
\begin{equation}
r_N^\mu(t) 
   = \frac{1}{\sqrt{N}}\sum_i \xi_i^\mu \tilde \sigma^\mu_i(t)
       + \sum_s 
           \frac{1}{N}  
	   \sum_i
	   \frac{\partial \sigma_i(t)}
                 {\partial \theta_i(s)}\Bigg|_{\theta_i(s)=0}         
          r_N^\mu(s)
       + \order{1/\sqrt{N}}
\end{equation}
by using the definition of $r^\mu_N(t)$.  We now take the limit
$N \to \infty$.  By construction, the first term on
the r.h.s. is a sum of independent terms and thus converges to a normal
distribution with zero mean and variance 1.  To the second term we apply the
LLN.  The result of the limit reads 
\begin{equation}
r^\mu(t) 
   =  \tilde r^\mu(t)
       + \sum_s 
          G(t,s)
          r^\mu(s) \, .
\end{equation}

This recursion relation for $r^\mu(s)$ corresponds to 
(\ref{eq:etat}) by using the  substitution
$r^\mu$  for $\eta$ and $\tilde r^\mu$ for $\tn$ as in (\ref{eq:bla}).
Finally, we  insert this relation into the expression (\ref{eq:h:recur}) of
the local field  at time $t+1$ leading to, after some algebra, the
expression for the exact local field as in equation
(\ref{eq:h:recursion}) for the GFA.
This shows that we have found all  feedback correlations.

\section{Other architectures}

In this section we shortly discuss the application of the SNA approach
to other architectures.

For the extremely diluted symmetric architecture one has found an exact
solution for the dynamics up to time step $3$ by using a probabilistic
method analogous to the SNA \cite{ZP92} and comparing it with the
generating functional approach \cite{WS91}.
In this case, the effective local field, starting from the GFA is given
by
\begin{equation}
    \label{eq:h:gen}
h(t) = \xi\,m(t) + \alpha \sum_{s=0}^{t-1} R(t,s) \, \sigma(s)
        + \sqrt{\alpha} \nn(t)
\end{equation}
with $\eta(t)$ a temporally correlated noise with zero mean and
correlation matrix $\bD$. Hereby,
\begin{equation}
\bD = \bC
  \qquad \qquad
\bR = \bG \, .
\end{equation}
It is clear that, besides the simplification in both the correlations and
retarded self-interaction, feedback correlations of arbitrary length
survive and  make the dynamics as hard to solve as the one of the fully
connected model. Hence, a completely analogous discussion as the one
before can be made in this case.

For the extremely diluted asymmetric architecture the effective local field
in the GFA approach is given by (\ref{eq:h:gen}), where now
\begin{equation}
\bD = \bC
  \qquad \qquad
\bR = 0 \,.
\end{equation}
Hence, the retarded self-interaction is zero, telling us
that the local field at time $t$ does not directly depend on the spins at
previous times but only indirectly via the noise.  This is precisely the
reason why for $\alpha=0$, the short-memory approximation gives the exact
answer (see above). It amounts to having a response function of the form
(\ref{eq:G:alpha=0}). Therefore, the SNA describes the correct dynamics.
The reason is that for asymmetric dilution the probability to have a loop of
finite length tends to zero in the thermodynamic limit \cite{DGZ87,
KZ91}.

For sequence processing networks the GFA effective local field is given by
(\ref{eq:h:gen}) with
\begin{equation}
\bD = \sum_{n\ge 0} (\bG^\dagger)^n \bC (\bG)^n
  \qquad \qquad
\bR = 0
\end{equation}
and the situation is analogous to the one for the asymmetrically diluted
model in the sense that the retarded self-interaction is  zero.  Again
we have a response function of the form (\ref{eq:G:alpha=0}) and, hence, the
short-memory approximation is exact. This is consistent with and further
explains the results in \cite{KO02}
where it has been shown through explicit calculation that
the order parameter equations obtained through the GFA are
equivalent to those of  statistical neurodynamics.

\section{Concluding remarks}

In this paper we have revisited the signal-to-noise approach for solving
the dynamics of the fully connected Little-Hopfield model by comparing it
with the exact generating functional analysis. In order to do so we have
derived a recursion relation for the effective local field in the
generating functional approach. We have shown that the signal-to-noise
analysis is a short-memory approximation that is exact up to the third
time step. For further time steps, it stays very accurate in the retrieval
region but not in the spin-glass region. These results are confirmed by
numerical simulations. The application of these methods to other
architectures and to sequence processing models has also been discussed.

\section*{Acknowledgments}

We would like to thank T. Coolen and G.M. Shim for informative
discussions. This work has been supported in part by the
Fund of scientific Research, Flanders-Belgium.


\begin{thebibliography}{99}

\bibitem{GCS03} Galla T, Coolen A C C and Sherrington D 2003 {\it
Preprint} cond-mat/0303615

\bibitem{BBSV03} Boll\'e D, Busquets Blanco J, Shim G M and Verbeiren T
 2003 {\it Preprint} cond-mat/0304553 (Physica A at Press)

\bibitem{C01} Coolen A C C 2001 in {\it Handbook of Biological Physics
Vol 4}, ed. by Moss F and Gielen S (Elsevier Science) 597

\bibitem{KO02} Kawamura M and Okada M 2002
 J. Phys. A: Math. Gen. {\bf  35} 253


\bibitem{MSR73} Siggia E D, Martin P C  and  Rose H A 1973
Phys. Rev. A, {\bf 8}423

\bibitem{Dom78} De Dominicis C 1978
J. Phys. A: Math. Gen. {\bf 18} 4913

\bibitem{RSZ89} Rieger H, Schreckenberg M and Zittartz J 1989
Z. Phys. B {\bf 74} 527

\bibitem{HBFKS89} Horner H, Bormann D, Frick M, Kinzelbach H and
Schmidt A 1989  Z. Phys. B {\bf 76} 381

\bibitem{AM88} Amari S and Maginu K 1988 Neural Networks {\bf 1} 63

\bibitem{O95} Okada M 1995 Neural Networks {\bf 8} 833


\bibitem{Nbook} Nishimori H 2001 {\it Statistical Physics of Spin
Glasses and Information Processing} (Oxford Univ. Press)

\bibitem{B03} Boll\'e D 2003 cond-mat/0307104 to appear in {\it Advances
in Condensed Matter and Statistical Mechanics} ed. by Korutcheva F and
Cuerno R (Nova Science Publishers)

\bibitem{K85} Kinzel W 1985 Z. Phys. B {\bf 60} 205

\bibitem{PZ91} Patrick A E and Zagrebnov V A  1991
   J. Phys. A: Math. Gen. {\bf 24} 3413

\bibitem{PZ91b}Patrick A E and Zagrebnov V A  1991
   J. Stat. Phys. {\bf 63} 59

\bibitem{P84} Peretto P 1984 Biol. Cybern. {\bf 50} 51

\bibitem{FK88} Fontanari J F and K\"oberle R 1988
   J. Physique {\bf 49} 13

\bibitem{DGZ87} Derrida B Gardner E and  Zippelius A 1987
   Europhys. Lett. {\bf 4} 167

\bibitem{KZ91} Kree R and Zippelius A 1991 in {\it Models of neural networks},
eds. Domany E van Hemmen J L and Schulten K (Springer, Berlin)

\bibitem{NO93} Nishimori H and Ozeki T 1993
   J. Phys. A: Math. Gen. {\bf 26} 859

\bibitem{HO90} Henkel R D and Opper M 1990
   Europhys. Lett. {\bf 11} 403

\bibitem{PZ90} Patrick A E and Zagrebnov V A 1990
 J. Phys France {\bf 51} 1129

\bibitem{BJS98} Boll{\'e} D, Jongen G and  Shim G M 1998
 J. Stat. Phys. {\bf 91} 125

\bibitem{HJ85} Horn R A and Johnson C R 1985
 {\em Matrix Analysis}, Cambridge University Press

\bibitem{BKS90} Barkai E, Kanter I and Sompolinsky H 1990
  Phys. Rev. A {\bf 41} 590

\bibitem{EO92} Eissfeller H and Opper M 1992
  Phys. Rev. Lett. {\bf 68} 2094


\bibitem{ZP92} Patrick A E and Zagrebnov V A 1992
  J. Phys A: Math. Gen. {\bf 25} 1009


\bibitem{WS91} Watkin T L H and Sherrington D 1991
  J. Phys A: Math. Gen. {\bf 24} 5427


\end{thebibliography}
\end{document}